# Liquid Mirror Telescopes: A progress report

Ermanno F. Borra, Marc Ferrari, Luc Girard, Gilberto Moretto, Grégoire Tremblay

Centre d'Optique, Photonique et Laser, Département de Physique, Université Laval, Québec, Qc, Canada G1K7P4

and Gérard Lemaître,

Observatoire de Marseille, 2 place LeVerrier, 13248, Marseille, cedex, France

## ABSTRACT

We review the present status of liquid mirror telescopes. Interferometric tests of liquid mirrors (the largest one having a diameter of 2.5 meters) show excellent optical qualities. The basic technology is now sufficiently reliable that it can be put to work. Indeed, a handful of liquid mirrors have now been built that are used for scientific work. A 3.7-m diameter LMT is presently being built in the new Laval upgraded testing facilities. Construction of the mirror can be followed on the Web site: http://astrosun.phy.ulaval.ca/lmt/lmt-home.html. Finally we address the issue of the field accessible to LMTs equipped with novel optical correctors. Optical design work, and some exploratory laboratory work, indicate that a single LMT should be able to access, with excellent images, small regions anywhere inside fields as large as 45 degrees.

## 1. INTRODUCTION

It has been known for centuries[1] that the surface of a spinning liquid takes the shape of a paraboloid that can be used as the primary mirror of a telescope. Following the suggestion[2] that modern technology now gives us tracking techniques that render liquid mirrors useful to Astronomy, a research and development program was begun to assess the feasibility of the concept, leading to the demonstration of a diffraction limited 1.5-m mirror[3], an article that also gives a wealth of technological details, followed by a 2.5-m mirror[4]. For imagery, narrow-band filter spectroscopy or slitless spectroscopy, one can use a technique, called time delayed integration and abbreviated as TDI, that uses a CCD detector that tracks by electronically stepping its pixels. The information is stored on disk and the nightly observations can be coadded with a computer to give long integration times. The technique has been demonstrated [5] with a 2.7-m diameter liquid mirror telescope. They show a deep sky exposure taken with that telescope.

Liquid mirror telescopes cannot be tilted, hence the issue of the field of view accessible to a liquid mirror is an important one, for the usefulness of a LMT increases greatly with its field of regard. We have begun investigating this topic, finding a corrector design that gives a very large field of regard.

At the time of this writing several liquid mirrors and liquid mirror telescopes have been built. This paper reports on the present status of liquid mirrors and liquid mirror telescopes.

## 2. LIQUID MIRRORS

One can easily show [2] that the surface of a rotating liquid takes the shape of a parabola. Using a reflecting liquid one gets a reflecting parabola, the ideal surface for the primary mirror of a telescope imaging an object placed at infinity. The focal length of the mirror L is related to the acceleration of gravity g and the angular velocity of the turntable ω by

$$L = g/(2\omega^2) \qquad (1)$$

For large mirrors of practical interest the periods of rotation are of the order of several seconds and the linear velocities at the rims of the mirrors range between 2 and 10 km/h. An exploded view of the basic mirror setup can be seen in Figure 1.

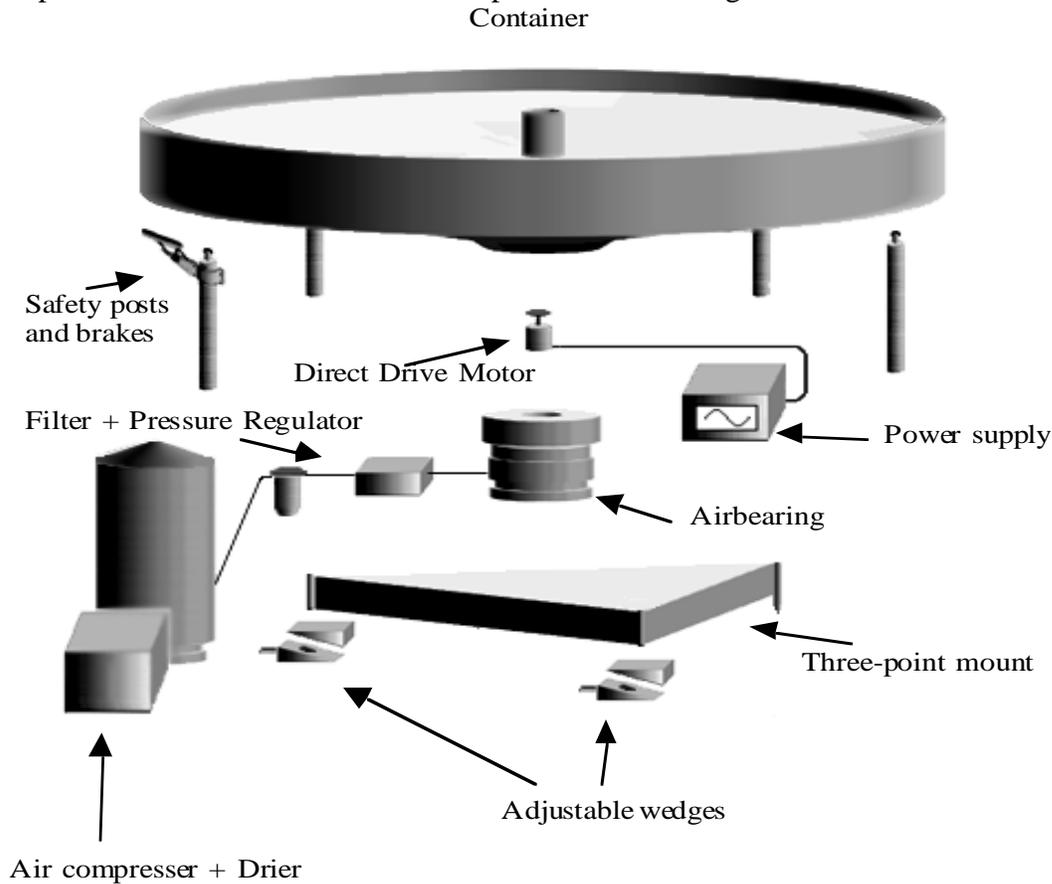

Figure 1. It shows an exploded view of the basic mirror setup.

The container and bearing are attached to a three-point mount that aligns the axis of rotation parallel to the gravitational field of the Earth. Alignment can be done with a spirit level to within one arcsecond, sufficient for many applications. More precise leveling devices exist. There are optical methods that can align it to greater accuracy.

All existing liquid mirrors use airbearings because they are convenient for small systems and commercially available units have the required precision and low friction. Oil bearings should be more advantageous for large mirrors since they can support substantially higher masses for a given bearing size. The turntable is driven by a synchronous motor. The motor is controlled by a variable-frequency AC power supply stabilized with a crystal oscillator. The container is a very important component of the system. It must be light, yet be rigid. Our latest containers are made of Kevlar laminated over a foam core. It is very important to work with thin mercury layers to minimize weight and, especially, to dampen disturbances. For this, we have developed techniques that allow us to work with layers of mercury as thin as 0.5 mm [3].

Table 1. Costs of components and labor needed to construct a 3.7-m mirror.

| Item | Cost of components (1996 US $) | Labor ($45/hour) |
|---|---|---|
| Mirror+accessories[1] | 32,000 | 22,000 |
| Safety equipment[2] | 4,000 | 6,00 |
| Installation[3] | 1,000 | 4,000 |
| Total | 37,000 | 32,000 |

[1] Complete system, including base, mercury, motor, etc...
[2] Includes mercury sniffer, safety brakes and anti-spill wheels
[3] In-situ installation, including balancing, debugging, checking image quality . Based on experience with the UWO 2.65-m LM [6]

A 3.7-m diameter LMT is presently being built in the new Laval upgraded testing facilities. Construction of the mirror can be followed on the Web site: http://astrosun.phy.ulaval.ca/lmt/lmt-home.html. Table 1 shows an estimate of the cost of building the 3.7-m mirror. At the time of this writing this mirror is only half built so that there is some extrapolation of the time needed for construction. It must be noted that this is a prototype and that a better engineered system would certainly be less expensive. It must also be noted that these are rough estimates. On the one hand there are hidden costs that are not accounted for (e.g. rent for construction space), on the other the construction time can be decreased given our experience building the present containers.

We have extensively tested a 2.5-m f/1.2 LM. Testing is done at the center of curvature and we therefore must use null lenses to correct the large spherical aberration present at the center of curvature of any parabolic mirror. The interferometry is done with scatter plate and Shack-cube interferometers. The interferograms are captured with 1/60 second exposure times by a 512X480 CCD detector connected to an 8-bit framegrabber interfaced to a microcomputer.

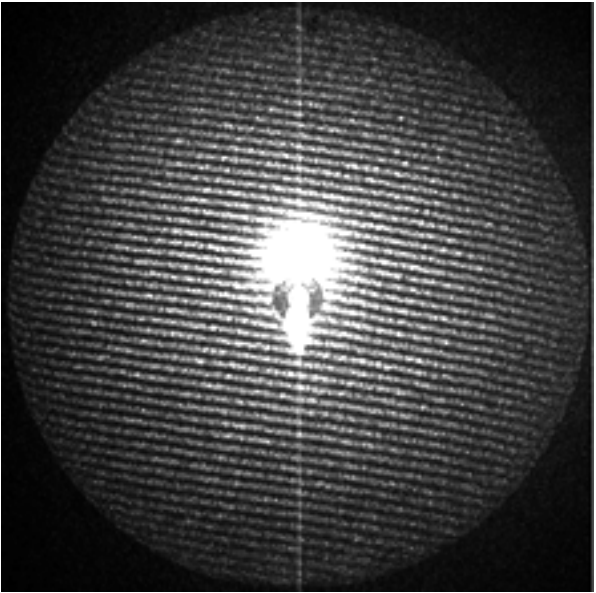

Figure 2. It shows a typical interferogram of the 2.5-m mirror.

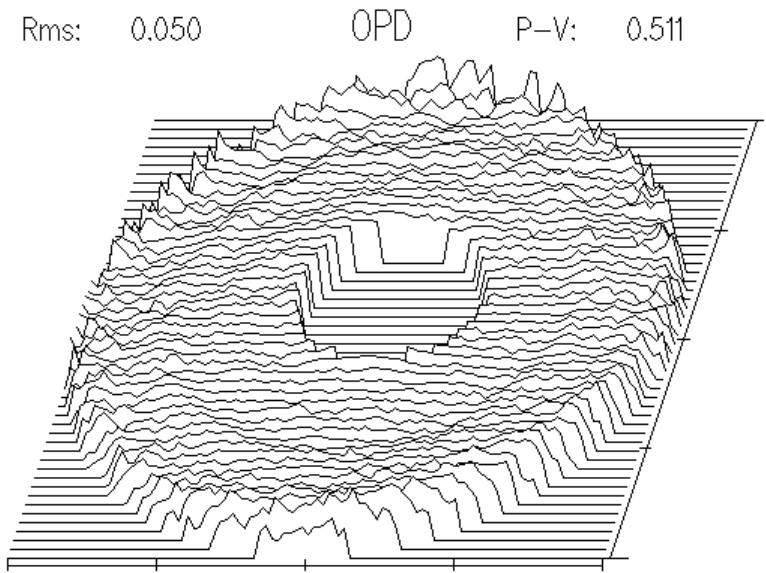

Figure 3. It shows the three-dimensional rendering of the surface of the 2.5-m mirror.

Figure 2 shows a typical interferogram of the 2.5-m mirror and Figure 3 shows the three-dimensional rendering of its surface. The statistics associated with the surface are given in units of surface deviations on the mirror at a wavelength of 6328 Å. The spatial resolution of the interferometry on the mirror is 3 cm. The 1/60 second capture times are sufficiently short that we can detect rapid liquid movements, but they also render the interferometry sensitive to seeing effects in the testing tower. Our testing facility is in a basement room lined with thick concrete walls that has small temperature gradients, so that seeing effects are minimal. Because the mirror is liquid, a few selected interferograms are not necessarily representative of its optical quality. We have analyzed a large number of interferograms and have videotaped hours of interferogram data to satisfy ourselves that the interferogram shown in Figure 2, and the wavefront of Figure 3 are representative.

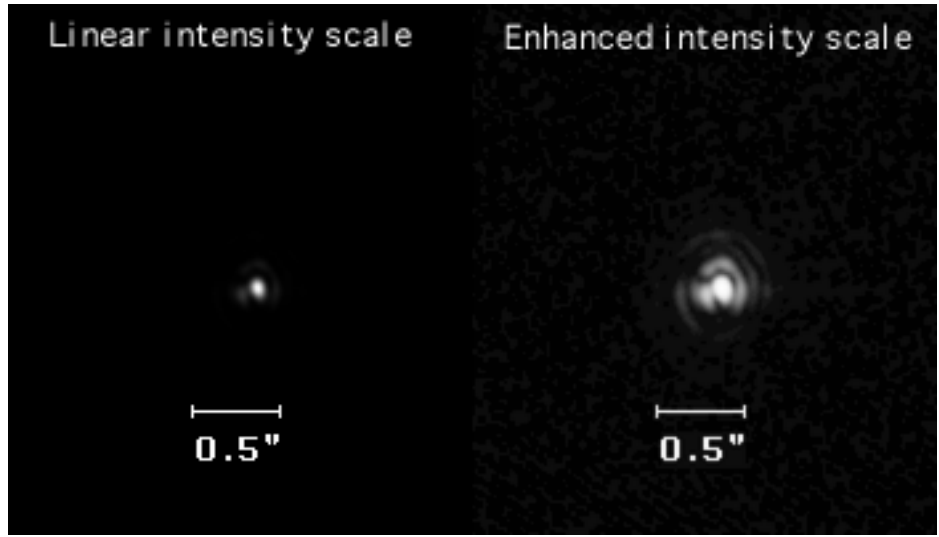

Figure 4. It shows the image of an artificial star imaged, with the 2.5-m mirror.

Figure 4 shows the image of an artificial star imaged, through null lenses, at the center of curvature of the 2.5-m mirror. We can see the Airy diffraction pattern of the 2.5-meter diameter liquid mirror. Fainter rings are present further away from the center of the PSF but are below detection in this image. We have videotaped hours of data and find that the Airy pattern is always visible, although the intensities and symmetries of the rings vary a little, probably from seeing in the testing tower. The observation of the Airy pattern (taken through the null lenses) gives direct evidence that the mirror is near diffraction limit.

The interferometry cannot detect defects smaller than its spatial resolution on the mirror. High spatial frequency ripples cause extended wings and scattered light far from the core of the PSF. Liquid mirrors are sensitive to vibrations but they are only a minor nuisance. Our mirrors are located in the basement of a large building that vibrates like any building does. We do see the effect of vibrations in the form of concentric rings on the surface of the mirror. However, their amplitudes are negligible ( ~/100 $\lambda$) for thin mercury layers. We can walk around our mirrors without inducing excessive vibrations.

We have made detailed analysis of the scattered light of liquid mirrors. We find that liquid mirrors do not have an undue amount of scattered light. To minimize scattered light, one must work with a thin layer of mercury to dampen vibration and wind effects. We also found that misalignments of the axis of rotation of the mirror introduce scattered light. Ideally, the axis should be vertical within 0.25 arcseconds; although errors below 1 arcseconds introduce tolerable levels of scattered light.

# 3. FIELD CORRECTORS FOR LIQUID MIRROR TELESCOPES

One of the often cited limitations of liquid mirror telescopes pertains to the small region of sky which they can observe. Because the aberrations of a parabola increase rapidly with field angle, classical corrector designs cannot yield subarcsecond images for angles significantly greater than one degree. To access larger fields, innovative corrector designs must be explored. A landmark design, using a fixed primary mirror, was given by Richardson & Morbey [7] who devised a three-mirror system that permits a 10-meters, f/5 LMT to be operated at 7.5 Deg from the zenith. Borra [8] has explored analytically how far off axis one can use a liquid mirror telescope and has shown that, in principle, aberrations can be corrected for zenith distances as large as 45 Deg.

A first exploration Borra, Moretto and Wang [9], investigated a two-mirror corrector used with a parabolic primary mirror observing at large zenith angles. The design, dubbed BMW corrector, uses complex surfaces that give subarcsecond images as far as 22.5 Deg from the optical axis. However, there then arises the practical problem of making the required complex anamorphic aspheric surfaces with existing technology .

As a further step toward a design feasible with existing technology, Moretto and Borra [10] further investigate the BMW design assuming mirrors that can be made with the Active Vase Mirror technology pioneered by Lemaître[11,12]. Members of our team[13] have tested a stainless steel AISI 420 prototype mirror having a 16 cm aperture, a controlled pressure load, and two series of 12-punctual radial force application positions distributed symmetrically in two concentric rings around the mirror.

Pure modes Coma3, Astm3 and Comatri are generated by the modulation of forces at the 12-punctual radial support positions. The pure modes and the superposition of these terms are shown in Figure 5.

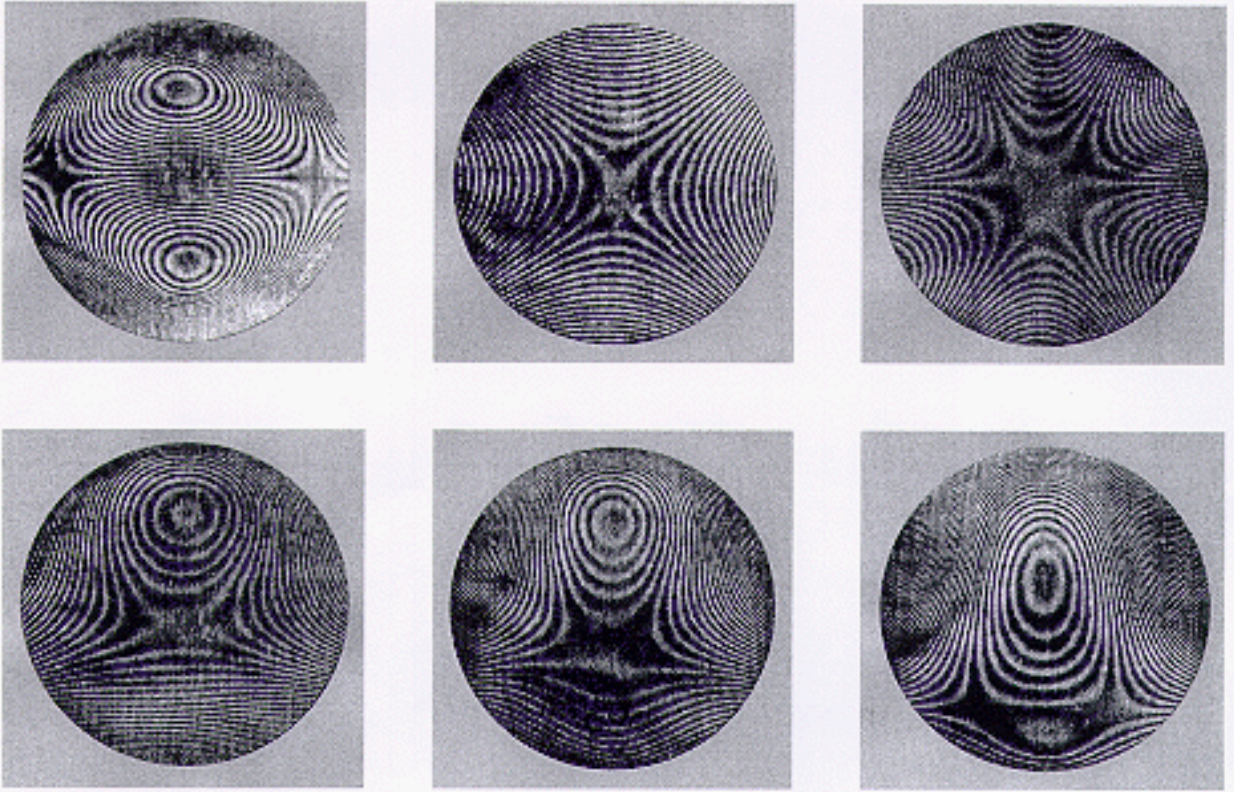

Figure 5. It shows the interferograms produced by the active vase mirror[13]. In the first row one can see the pure deformation modes: from left to right Coma3, Astm3 and Comatri. In the second row the superposed modes, from left to right:
Tilt + Astm3 + Coma3, Tilt + Astm3 + Coma3 + Comatri, Tilt + Astm3 + Coma3 + Comatri + Sph3

Moretto and Borra[10] have explored BMW correctors design for a 4.0 m f/5.25 Liquid Mirror observing at 7.5 Deg from the zenith having a field of view of 18 Arcmin and having two vase mirrors as secondary and tertiary mirrors warped with Zernike polynomials. All of the polynomials terms used in this design were demonstrated experimentally in [13]. The optimizations were done for the wavelength band 6000 to 8000 Å. The geometrical parameters, in millimeters, are given in Figure 6. To better control the distortion we introduce, after the tertiary mirror, three additional lenses: the largest one having spherical surfaces and the other two aspherical surfaces.

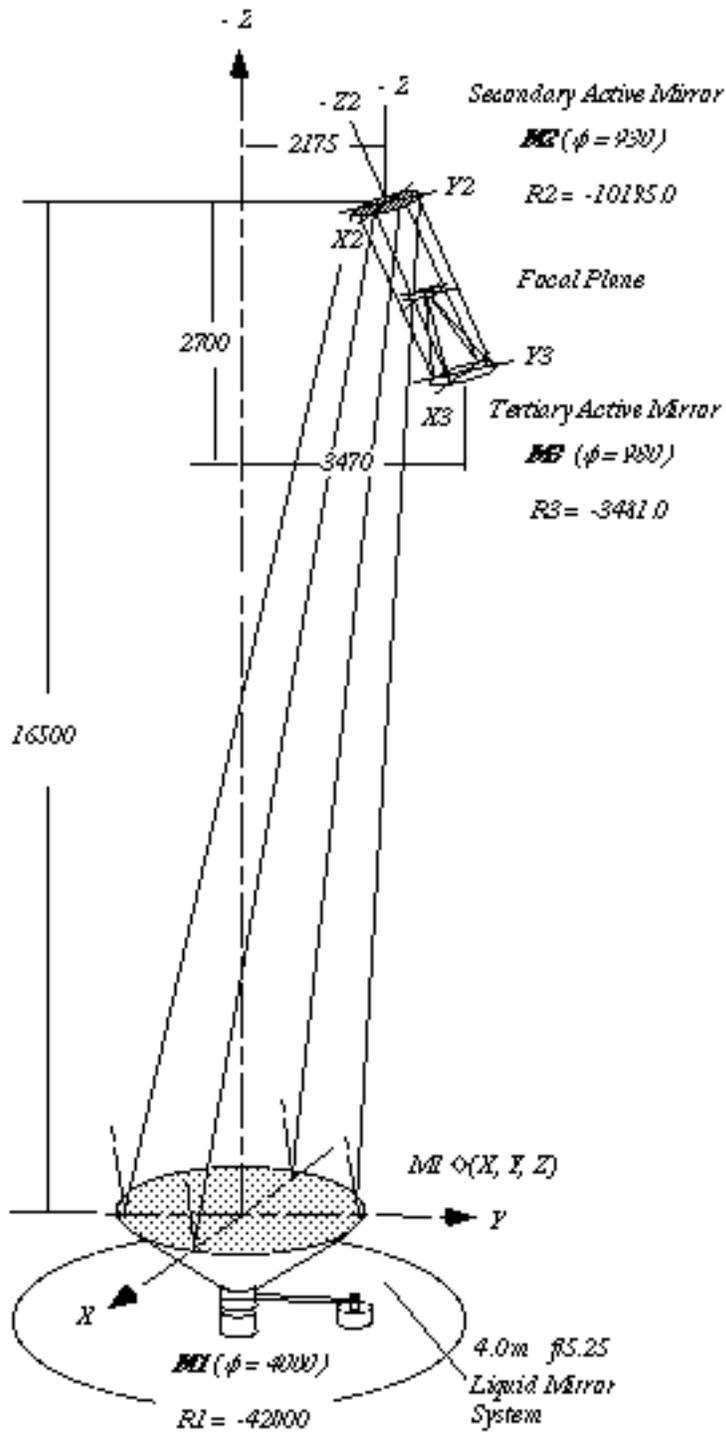

Figure 6. This diagram shows a schematic design for a 4-m diameter f/5.25 mirror with the correctors observing at θ = 7.50 Deg from the zenith and with a field of view of 18 Arcmin. The parameter φ represents the diameter for each mirror and Rn is the radius of curvature for the n-th surface. The dimensions are given in millimeters. Note that the system is symmetrical in X-Z plane.

The radial energy distribution of the PSF for each field in terms of the 50 %, 80 % and 100 % encircled energy diameters (EED) and the RMS spot diameter (RMS-SD) are shown in Table 2.

Table 2. Radial energy distribution for the PSF which characterizes the corrections at the selected zenith angles: $\theta = 7.50$ +/- 0.15 Deg. The optimization was done for the wavelength band 6000 - 8000 Å.

| Field(x,y) [Deg] | RMS-SD [Arcseconds] | EED [Arcseconds] 50 % | 80% | 100% |
|---|---|---|---|---|
| -0.15, 7.35) | 0.771 | 0.502 | 0.941 | 1.870 |
| (-0.15, 7.50) | 0.663 | 0.497 | 0.834 | 1.783 |
| (-0.15, 7.65) | 0.777 | 0.556 | 0.991 | 1.861 |
| (0.00, 7.35) | 0.580 | 0.384 | 0.814 | 1.224 |
| (0.00, 7.50) | 0.628 | 0.454 | 0.781 | 1.671 |
| (0.00, 7.65) | 0.699 | 0.409 | 0.999 | 1.453 |
| (+0.15, 7.35) | 0.771 | 0.502 | 0.941 | 1.870 |
| (+0.15, 7.50) | 0.663 | 0.497 | 0.834 | 1.783 |
| (+0.15, 7.65) | 0.777 | 0.556 | 0.991 | 1.861 |

In principle the telescope could track by moving and warping the mirrors to follow an object in the sky, but the geometry of the corrector and the shapes of the mirrors must change in real time to track. A simpler system can use a rigid corrector with the TDI technique (see the introduction). The optical and mechanical setups are then simple since the corrector is set for a particular zenith distance and does not have to be adjusted to work at different zenith distances. A corrector designed for a given zenith distance (e.g. $\theta = 7.5$ degrees) can be used to observe objects passing anywhere within a field of view of $2\theta$ (e.g. 15 degrees) by moving it at different azimuths but at a fixed zenith distance. This is the only degree of freedom allowed, everything else (e.g. mirror shapes and distances) are fixed.

Once a sufficiently large accessible field is achieved, a fixed primary and movable correctors yield a more efficient system than a classical tiltable telescope. A classical telescope can only observe a field at a time, while a fixed primary with several correctors could access many widely separated fields simultaneously. One can readily envision a corrector tracking a field North of the zenith at the same time as another one tracks in the South, while fixed correctors at the edge of the field carry out surveys in the driftscanning mode. This primary-sharing setup, allowing several research programs to be carried out simultaneously, is particularly attractive for very large telescopes, for which observing time is at a premium.

# 4. SOME PRESENT APPLICATIONS OF LIQUID MIRRORS

Because optics are so important in scientific, medical and engineering instrumentation, inexpensive large high-quality liquid mirrors are bound to have an impact in those fields. At the time of this writing several liquid mirrors and liquid mirror telescopes are at work in a variety of fields of Science and Engineering. As mentioned in the Introduction and [5], a 2.7-m LMT has been used for an astronomical survey. A 5-m LMT that will continue the survey is presently under construction.

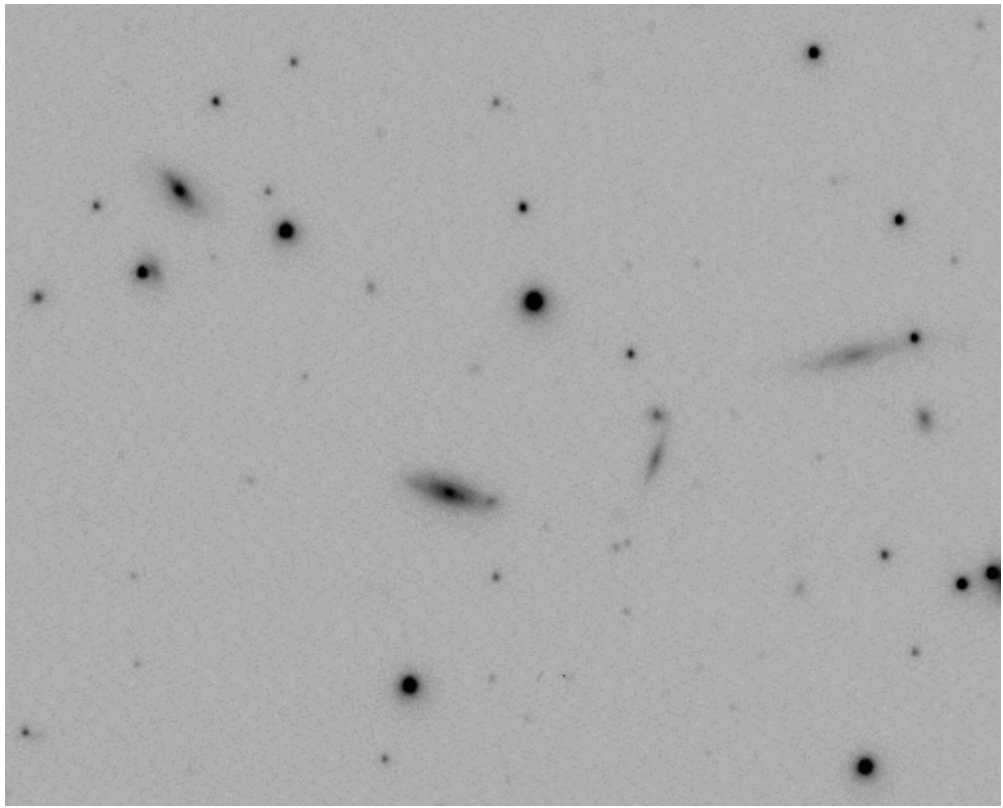

Figure 5. It shows a CCD image of a region of sky 5 arcminutes long observed with a 3-m liquid mirror telescope built and operated by NASA. The image is a courtesy of Mark Mulrooney.

Figure 5 shows a CCD image of a region of sky 7 arcminutes long observed with a 3-m liquid mirror telescope built and operated by NASA. It was obtained from a single nightly pass that gives a 100-second exposure. It reaches about magnitude 23 and shows several faint galaxies and stars. Amospheric scientists have expressed interest for these inexpensive large mirrors for lidar applications: A lidar facility at the University of Western Ontario housing a 2.6-m diameter liquid mirror receiver has been built and is in operation[6]. UCLA has recently built a similar facility in Alaska. Liquid mirrors are also used to generate reference surfaces to test optical surfaces. The work of Ninane[14] is particularly of interest, not only because of her original use of a liquid mirror, but also because she carried out independently optical shop tests that confirm the high optical quality of liquid mirrors reported in[3] and[4]. It also worthwhile to point out an interesting application of active flat mercury mirrors[15].

## 5. CONCLUSION

Interferometric tests of liquid mirrors having diameters as large as 2.5 meters show excellent optical qualities. Although there is room for improvement of this very young technology, we do have a working design that is sufficiently robust to be useful for practical use. We also have an adequate understanding of the behavior of a liquid mirror under perturbations. In other words, we know how to make liquid mirrors that work but one can do better.

A handful of liquid mirrors have now been built and are used for scientific work: astronomy, atmospheric sciences, space sciences, optical shop tests. More applications are certainly forthcoming given the advantages of liquid mirrors, foremost of which is cost.

The issue of the field accessible to a LMT is a very important one since the usefulness of a LMT increases with its accessible field. Optical design work, indicates that a single LMT should be able to access fields as large as 45 degrees. We only are at the beginning of this exploratory work and additional effort can probably find simpler systems with improved performance. On the other hand, those correctors can only increase the cost of a LMT and they can only be practical if the total instrument is sufficiently cheaper than a conventional telescope.

## 6. ACKNOWLEDGMENTS

This research has been supported by Natural Sciences and Engineering Research Council of Canada and Formation des Chercheurs et Aide à la Recherche grants. We wish to thank Mark Mulrooney for taking and sending us the image used to generate Figure 5.